\documentclass[ALICE,manyauthors]{cernphprep}

\usepackage[comma,square,numbers,sort&compress]{natbib}
\usepackage{hyperref}
\usepackage{color}
\usepackage{lineno}
\usepackage{booktabs}
\usepackage{xspace}

\newcommand{\pt}{\ensuremath{p_{\mathrm T}}} 
\newcommand{\snn}{\mbox{$\sqrt{s_{\mathrm{NN}}}$}}   
\newcommand{\pb}{Pb--Pb }
\newcommand{\ppb}{p--Pb }

\newcommand{\pp}{\mbox{pp}}

\newcommand{\jpsi}{\mbox{J/$\psi$ }}
\newcommand{\psip}{$\psi$(2S)}

\newcommand{\dndeta}{\ensuremath{{\rm d}N_{\rm ch}/{\rm d}\eta}\xspace}
\newcommand{\dndetar}{\ensuremath{{\rm d}N_{\rm ch}^{R}/{\rm d}\eta}\xspace}
\newcommand{\ntr}{\ensuremath{N_{\mathrm{trk}}}\xspace}
\newcommand{\ntrcorr}{\ensuremath{N_{\mathrm{trk}}^{\mathrm{corr}}}\xspace}
\newcommand{\nch}{\ensuremath{N_{\mathrm{ch}}}\xspace}

\newcommand{\mpt}{\ensuremath{\langle {p}_{\mathrm T} \rangle}\xspace}
\newcommand{\rpb}{\mbox{$R_{\mathrm{pPb}}$}}
\newcommand{\acef}{$A\! \! \! \times \! \! \varepsilon$}
\newcommand\y{$y$}
\newcommand{\ycms}{\ensuremath{y_{\mathrm{cms}}}}
\newcommand{\etacms}{\mbox{$\eta_{\mathrm{cms}}$}}
\newcommand\gevc{GeV/$c$}
\newcommand{\average}[1]{\ensuremath{\langle #1 \rangle}\xspace}

\newcommand{ \be }{\begin{eqnarray}}
\newcommand{ \ee }{\end{eqnarray}}

\newcommand{ \rms }[1]{r.m.s.}

\begin{document}%

%%%%%%%%%%%%%%%  Title page %%%%%%%%%%%%%%%%%%%%%%%%
\begin{titlepage}
\PHyear{2017}
\PHnumber{056}      % required, will be obtained from PH
\PHdate{30 March}  % required, will be obtained from PH
%

%%% Put your own title + short title here:
\title{\jpsi production as a function of charged-particle pseudorapidity density in \ppb collisions at \snn~ = 5.02 TeV}
\ShortTitle{\jpsi production as a function of charged-particle pseudorapidity density in \ppb}   % appears on right page headers

%%% Do not change the next lines
\Collaboration{ALICE Collaboration\thanks{See Appendix~\ref{app:collab} for the list of collaboration members}}
\ShortAuthor{ALICE Collaboration} % appears on left page headers, do not change

\begin{abstract}
We report measurements of the inclusive \jpsi yield and average transverse momentum as a function of charged-particle pseudorapidity density \dndeta in p--Pb collisions at \snn ~= 5.02 TeV with ALICE at the LHC. The observables are normalised to their corresponding averages in non-single diffractive events. An increase of the normalised \jpsi yield with normalised \dndeta, measured at mid-rapidity, is observed at mid-rapidity and backward rapidity.
%, similar to that found earlier in pp collisions. 
At forward rapidity, a saturation of the relative yield is observed for high charged-particle multiplicities. The normalised average transverse momentum at forward and backward rapidities increases with multiplicity at low multiplicities and saturates beyond moderate multiplicities. In addition, the forward-to-backward nuclear modification factor ratio is also reported, showing an increasing suppression of \jpsi production at forward rapidity with respect to backward rapidity for increasing charged-particle multiplicity.
\end{abstract}

\vspace{3cm}
%\centerline{Date: \today}

\end{titlepage}
\setcounter{page}{2}

%
%
%%%%%%%%%%% put the body of the article here
\section{Introduction}
\label{sec:intro}
%!TEX root = ../JpsiVsChMult.tex
 
Quarkonium states, such as the \jpsi meson, are prominent probes of the deconfined state of matter, the Quark-Gluon Plasma (QGP), formed in high-energy heavy-ion collisions \cite{Matsui:1986aa}. A suppression of \jpsi production in nucleus-nucleus (AA) collisions with respect to that in proton-proton (\pp) collisions has been observed by several experiments \cite{Alessandro:2005aa,Arnaldi:2007aa,Adare:2011aa,Abelev:2012ab,Adamczyk:2013tvk,Abelev:2014ad,Chatrchyan:2012aa,Aad:2011aa,Adam:2016rdg}. A remarkable feature is that, for \jpsi production at low transverse momentum (\pt) at the Large Hadron Collider (LHC), the suppression is significantly smaller than that at lower energies \cite{Abelev:2012ab,Abelev:2014ad,Adam:2016rdg}. The measurements of \jpsi production in proton (deuteron)-nucleus collisions, where the formation of the QGP is not expected, are essential to quantify effects (often denoted ``cold nuclear matter, CNM, effects''), present also in AA collisions but not associated to the QGP formation. At LHC energies, gluon shadowing/saturation is the most relevant effect which was expected to be quantified with measurements in \ppb collisions~\cite{Salgado:2012aa,Vogt:2013aa}. Furthermore, a novel effect, coherent energy loss in CNM (medium-induced gluon radiation), was proposed \cite{Arleo:2013aa}.

The measurements in d--Au collisions at the Relativistic Heavy Ion Collider (RHIC) have underlined the role of CNM effects in \jpsi production at $\snn=200$ GeV \cite{Adare:2008aa,Adare:2011ab,Adare:2013aa}. At the LHC, the first measurements of \jpsi production in minimum-bias \ppb collisions at $\snn=5.02$ TeV \cite{Aaij:2013zxa,Abelev:2014aa} showed that \jpsi production in \ppb collisions is suppressed at forward rapidity with respect to the expectation from a superposition of nucleon-nucleon collisions. The data have been further analysed to provide more differential measurements and discussed in comparison with several theoretical models \cite{Adam:2015ac}. A fair agreement is observed between data and models including nuclear shadowing \cite{Vogt:2013aa} or saturation \cite{Ma:2015sia,Ducloue:2015gfa}; also models including a contribution from coherent energy loss in CNM \cite{Arleo:2013aa} describe the data. These measurements are also relevant with respect to \jpsi production in \pb collisions at the LHC \cite{Abelev:2012ab,Abelev:2014ad}, currently understood to be strongly influenced by the presence of a deconfined medium. The measurements of $\Upsilon$ production in minimum-bias \ppb collisions at the LHC \cite{Aaij:2014ab,Abelev:2015ab} are also consistent with predictions based on CNM effects. Recent measurements of the $\psi$(2S) state in \ppb collisions have revealed a larger suppression than that measured for \jpsi production \cite{Abelev:2014ac,Aaij:2016eyl}. Such an observation was not expected from the available predictions based on CNM effects.

Concurrently, measurements of two-particle angular correlations in \ppb collisions at the LHC \cite{Chatrchyan:2013aa,Abelev:2013ad,Aad:2013aa,Abelev:2013ae,Chatrchyan:2013ab,Abelev:2014ae,Abelev:2015aa} revealed for high-multiplicity events features that, in \pb collisions, have been interpreted as a result of the collective expansion of a hot and dense medium. 
Furthermore, the identified particle \pt ~spectra \cite{Abelev:2013haa} show features akin to those in \pb collisions, where models  including collective flow, assuming local thermal equilibrium, agree with the data.

The measurement of \jpsi production as a function of centrality in \ppb collisions at the LHC \cite{Adam:2015jsa} showed that the nuclear effects depend on centrality.
$\Upsilon$ production has been studied as a function of charged-particle multiplicity in \pp ~and \ppb collisions by the CMS collaboration \cite{Chatrchyan:2014aa}. The yields of $\Upsilon$ mesons increase with multiplicity, while a decrease of the relative production of $\Upsilon(\mathrm{2S})$ and $\Upsilon(\mathrm{3S})$ with respect to $\Upsilon(\mathrm{1S})$ is observed.
The measurement of D-meson production as a function of event multiplicity in \ppb collisions \cite{Adam:2016mkz} exhibits features similar to those observed earlier in pp collisions, both for \jpsi \cite{Abelev:2012aa} and D-meson \cite{Adam:2015ota} production.

In this Letter measurements of the inclusive \jpsi yield and average transverse momentum as a function of charged-particle pseudorapidity density in \ppb collisions at $\snn=5.02$ TeV are presented. Performed in three ranges of rapidity for \pt ~$>0$ with the ALICE detector at the LHC, these measurements complement the studies of \jpsi and $\psi(2S)$ production as a function of the event centrality estimated from the energy deposited in the Zero Degree Calorimeters (ZDC) \cite{Adam:2015jsa,Abelev:ab}.
A measurement as a function of the charged-particle multiplicity does not require an interpretation of the event classes in terms of the collision geometry. 
Importantly, it enables the possibility to study rare events where collective-like effects may arise. 
The present data allow the investigation of events with very high multiplicities of charged particles, corresponding to less than 1\% of the hadronic cross section and establish as well a connection to the recent measurements of D-meson production as a function of event multiplicity \cite{Adam:2016mkz}. A measurement of the forward-to-backward \jpsi nuclear modification factor ratio is also presented.

\section{Experiment and data sample}
\label{sec:data}
%!TEX root = ../JpsiVsChMult.tex

The ALICE central barrel detectors are located in a solenoidal magnetic field of 0.5 T. The main tracking devices in this region are the Inner Tracking System (ITS), which consists of six layers of silicon detectors around the beam pipe, and the Time Projection Chamber (TPC), a large cylindrical gaseous detector providing tracking and particle identification via specific energy loss. Tracks are reconstructed in the active volume of the TPC within the pseudorapidity range $|\eta|<0.9$ in the laboratory frame.  The first two layers of the ITS ($|\eta| < $ 2.0 and $|\eta| < $ 1.4), the Silicon Pixel Detector (SPD), are used for the collision vertex determination and the charged-particle multiplicity measurement. The minimum-bias (MB) events are triggered requiring the coincidence of the two V0 scintillator arrays covering 2.8 $< \eta <$ 5.1 and $-3.7 < \!\eta \!< \!-1.7$, respectively. The two neutron Zero Degree Calorimeters (ZDC), placed at 112.5 m on both sides of the interaction point, are used to reject electromagnetic interactions and beam-induced background. 
The muon spectrometer, covering $-4 < \eta < -2.5$, consists of a front absorber, a 3 $\rm{T} \cdot \rm{m}$ dipole magnet, ten tracking layers, and four trigger layers located behind an iron-wall filter. In addition to the MB trigger condition, the dimuon trigger requires the presence of two opposite-sign particles in the muon trigger chambers.  The trigger comprises a minimum transverse momentum requirement of \pt~$> 0.5$ GeV/$c$ at track level.  The single-muon trigger efficiency curve is not sharp; the efficiency reaches a plateau value of $\sim 96\%$ at $\pt \sim$~1.5 GeV/$c$. The ALICE detector is described in more detail in~\cite{Abelev:2008aa} and its performance is outlined in \cite{Abelev:2014ffa}. 

The results presented in this Letter are obtained with data recorded in 2013 in \ppb collisions at \snn ~= 5.02 TeV. MB events are used for the \jpsi reconstruction in the dielectron channel at mid-rapidity. The dimuon-triggered data have been taken with two beam configurations, allowing the coverage of both forward and backward rapidity ranges.
In the period when the dimuon-triggered data sample was collected, the MB interaction rate reached a maximum of 200 kHz, corresponding to a maximum pile-up probability of about 3\%. The MB-triggered events used for the dielectron channel analysis were collected in one of the beam configurations at a lower interaction rate (about 10 kHz) and consequently had a smaller pile-up probability of 0.2\%.

Due to the asymmetry of the beam energy per nucleon in \ppb collisions at the LHC, the nucleon-nucleon center-of-mass rapidity frame is shifted in rapidity by $\Delta y =0.465$ with respect to the laboratory frame in the direction of the proton beam. This leads to a rapidity coverage in the nucleon-nucleon center-of-mass system $-1.37 < y_{\rm{cms}} < 0.43$ for the MB events, while the coverage for the dimuon-triggered data for the two different beam configurations is $-4.46 < y_{\rm{cms}} <$ $-2.96$ (muon spectrometer located in the Pb-going direction) 
and 2.03 $< y_{\rm{cms}} <$ 3.53 (muon spectrometer located in the p-going direction). The integrated luminosities used in this analysis are 51.4$\pm$1.9 $\mu$b$^{-1}$ (mid-rapidity), 5.01$\pm$0.19 nb$^{-1}$ (forward $y$) and 5.81$\pm$0.20 nb$^{-1}$ (backward $y$).

\section{Charged-particle pseudorapidity density measurement}
\label{sec:mult}
%!TEX root = ../JpsiVsChMult.tex

The charged-particle pseudorapidity density \dndeta is measured at midrapidity, $|\eta|<1$, and is based on the SPD information. Tracklets, i.e. track segments built from hit pairs in the two SPD layers, are used together with the interaction vertex position, which is also determined with the SPD information \cite{Abelev:2013ab}. Several quality criteria are applied to select only events with an accurate determination of the $z$ coordinate of the vertex, $z_{\mathrm{vtx}}$. To ensure full SPD acceptance for the tracklet multiplicity \ntr evaluation within $|\eta|<1$, the condition $|z_{\mathrm{vtx}}|< 10$ cm is applied for the selection of the events. 

During the data taking period about $8\%$ of the SPD channels were inactive, the exact fraction being time-dependent.
 The impact of the inactive channels of the SPD on the tracklet multiplicity measurement varies with $z_{\mathrm{vtx}}$. 
A $z_{\mathrm{vtx}}$-dependent correction factor is determined from data, as discussed in \cite{Abelev:2012aa}. This factor also takes into account the time-dependent variations of the fraction of inactive SPD channels.
The correction factor is randomised on an event-by-event basis using a Poisson distribution in order to emulate the dispersion between the true charged-particle multiplicity and the measured tracklet multiplicities.

The overall inefficiency, the production of secondary particles due to interactions in the detector material, particle decays and fake-tracklet reconstruction lead to a difference between the number of reconstructed tracklets and the true primary charged-particle multiplicity \nch ~(see details in \cite{Abelev:2013ab})\footnote{In this context, we regard as primary charged-particles all prompt charged particles including all decay products except products from weak decays of light flavour hadrons and of muons.}. 
Using  events simulated with the DPMJET event generator \cite{Roesler:2000he}, the correlation between the tracklet multiplicity (after the $z_{\mathrm{vtx}}$-correction), \ntrcorr, and the generated primary charged particles \nch ~is determined. 
The correction factor $\beta$ to obtain the average \dndeta value corresponding to a given \ntrcorr bin is computed from a linear fit of the \ntrcorr - \nch ~correlation. 
The charged-particle pseudorapidity density value in each multiplicity bin is given relative to the event-averaged value and is calculated as: 
$\dndetar = \dndeta / \average{\dndeta} = \beta \cdot \average{\ntrcorr} / \left(\Delta \eta \cdot \average{\dndeta}\right)$, where $\Delta\eta=2$ and \average{\dndeta} is the charged-particle pseudorapidity density for non-single diffractive (NSD) collisions, which was measured  to be $ \average{\dndeta} = 17.64 \pm 0.01 (\rm{stat.}) \pm 0.68 (\rm{syst.})$ %for \ppb collisions in $|\eta|<$ 1 
\cite{Abelev:2013ab}. The resulting values for the multiplicity bins are summarised in Tables~\ref{multiplicitypPb} and \ref{multiplicitypPbMidy} for forward and mid-rapidity, respectively.  
For the data at backward rapidity, the values are well within the uncertainties of those at forward rapidity.

 \begin{table}[h!]
\centering
\begin{tabular}{crrrr}
\toprule
\dndeta & \dndetar & $\sigma / \sigma_{MB}$\\
\midrule
4.8 $\pm$ 0.2 & 0.27 $\pm$ 0.01 & 26.4\% \\
10.9 $\pm$ 0.4 & 0.62 $\pm$ 0.03 & 14.3\% \\
14.6 $\pm$ 0.5 & 0.83 $\pm$ 0.03  & 7.9\% \\
17.8 $\pm$ 0.5 & 1.01 $\pm$ 0.04 & 9.6\% \\
21.4 $\pm$ 0.7 & 1.22 $\pm$ 0.05 & 8.5\% \\
25.0 $\pm$ 0.8 & 1.42 $\pm$ 0.06 & 7.2\% \\
28.6 $\pm$ 0.8 & 1.62 $\pm$ 0.06 & 6.0\% \\
32.7 $\pm$ 1.0 & 1.85 $\pm$ 0.07 & 6.7\% \\
37.8 $\pm$ 1.1 & 2.14 $\pm$ 0.08 & 4.6\% \\
44.2 $\pm$ 1.3 & 2.51 $\pm$ 0.10 & 4.2\% \\
54.3 $\pm$ 1.6 & 3.08 $\pm$ 0.12 & 2.4\% \\
71.4 $\pm$ 2.1 & 4.05 $\pm$ 0.16  & 0.3\% \\
\bottomrule
\end{tabular}
\caption{ \label{multiplicitypPb} Average charged-particle pseudorapidity density values (absolute and relative) in each multiplicity bin, obtained from \ntrcorr measured in the range $|\eta| <1$. The values correspond to the data sample used for the forward rapidity analysis. Only systematic uncertainties are shown since the statistical ones are negligible. The fraction of the MB cross section for each multiplicity bin is also indicated.}
\end{table}

\begin{table}[h!]
\centering
\begin{tabular}{crrrr}
\toprule
\dndeta & \dndetar & $\sigma / \sigma_{MB}$\\
\midrule
6.9 $\pm$ 0.2  & 0.39 $\pm$ 0.02 &  47.2\% \\
22.9 $\pm$ 0.6 & 1.30 $\pm$ 0.05 & 39.7\%  \\
42.3 $\pm$ 1.1 & 2.40 $\pm$ 0.10  & 10.9\%  \\
64.4  $\pm$ 1.6  &  3.65  $\pm$ 0.15  & 1.0\% \\
\bottomrule
\end{tabular}
\caption{ \label{multiplicitypPbMidy} 
As Table~\ref{multiplicitypPb}, but for the analysis of \jpsi production at mid-rapidity.
} 
\end{table}

The fraction of the MB cross section contained in each multiplicity bin ($\sigma / \sigma_{MB}$, derived from the respective event counts in the multiplicity bins and total number of MB events) is reported in Tables~\ref{multiplicitypPb} and \ref{multiplicitypPbMidy}. The softest MB events, which lead to absence of tracklets in $|\eta|<1$ are not accounted for in this analysis. They correspond to 
1.2\% of $\sigma_{MB}$ for the MB-triggered events and 1.9\% for the muon-triggered data; the difference is due to the different fraction of inactive channels in SPD and affects, albeit in a negligible way, only our first multiplicity bin.

The multiplicity selection in this analysis allows to sample the data in bins containing a small fraction of the MB cross section. Therefore, it gives the possibility to study the \jpsi production in rare high-multiplicity events which were not accessible in the centrality-based analysis \cite{Adam:2015jsa} (where the most-central event class corresponds to the range 2-10\% in $\sigma / \sigma_{MB}$).

\section[J/psi measurement]{\jpsi measurement}
\label{sec:meas}
%!TEX root = ../JpsiVsChMult.tex
 For the \jpsi analysis at forward and backward rapidities, muon candidates are selected by requiring the reconstructed track in the muon chambers to match a track segment in the trigger chambers. Furthermore, the radial distance of the muon tracks with respect to the beam axis at the end of the front absorber is required to be between 17.6 and 89.5 cm. This criterion rejects tracks crossing the high-density part of the absorber, where the scattering and energy-loss effects are large. A selection on the muon pseudorapidity $-4  < \eta <  -2.5 $ is also applied to reject muons at the edges of the spectrometer's acceptance. 

\begin{figure}[htb]
  {\centering 
\includegraphics[width=0.48\textwidth]{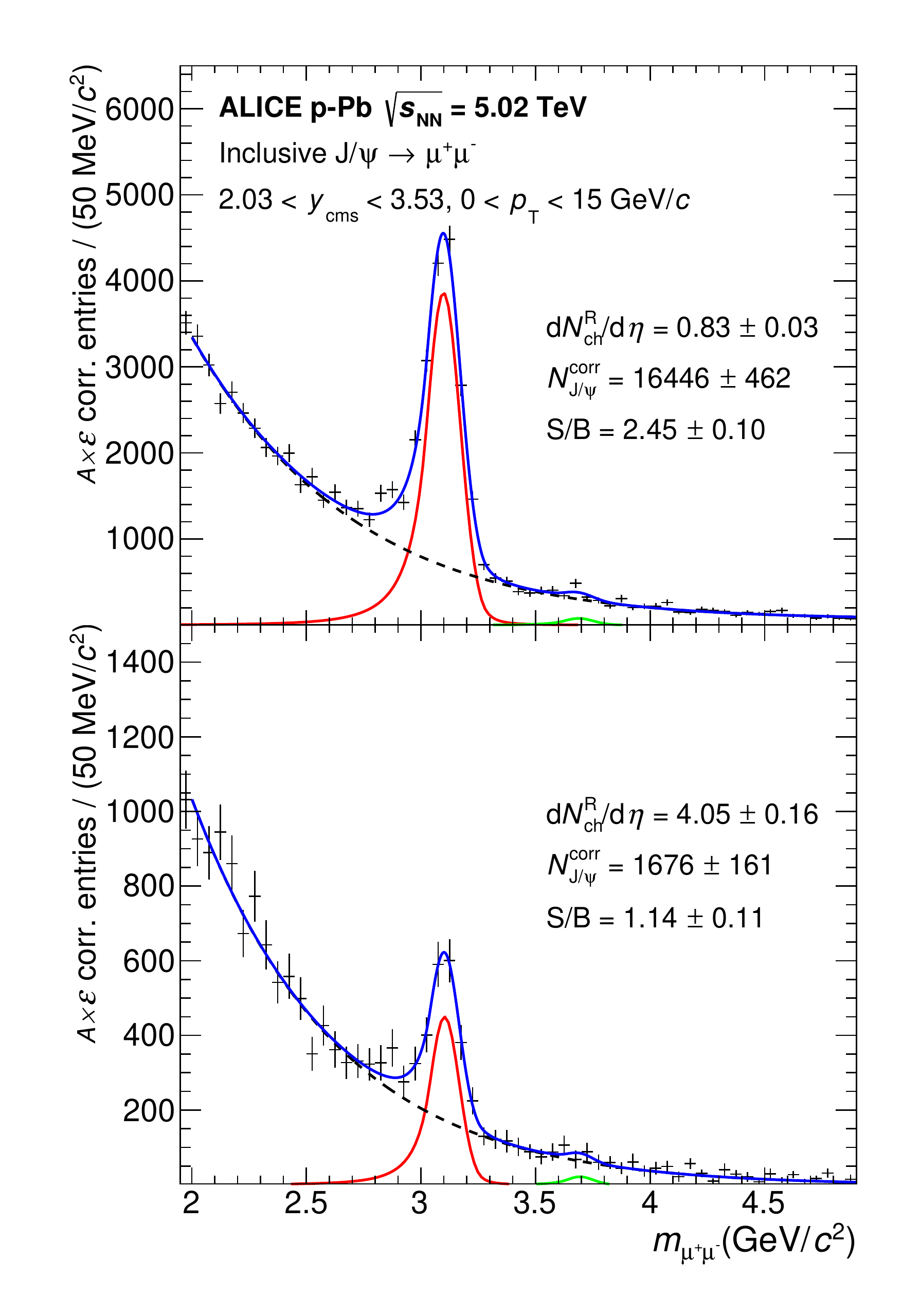}
\includegraphics[width=0.48\textwidth]{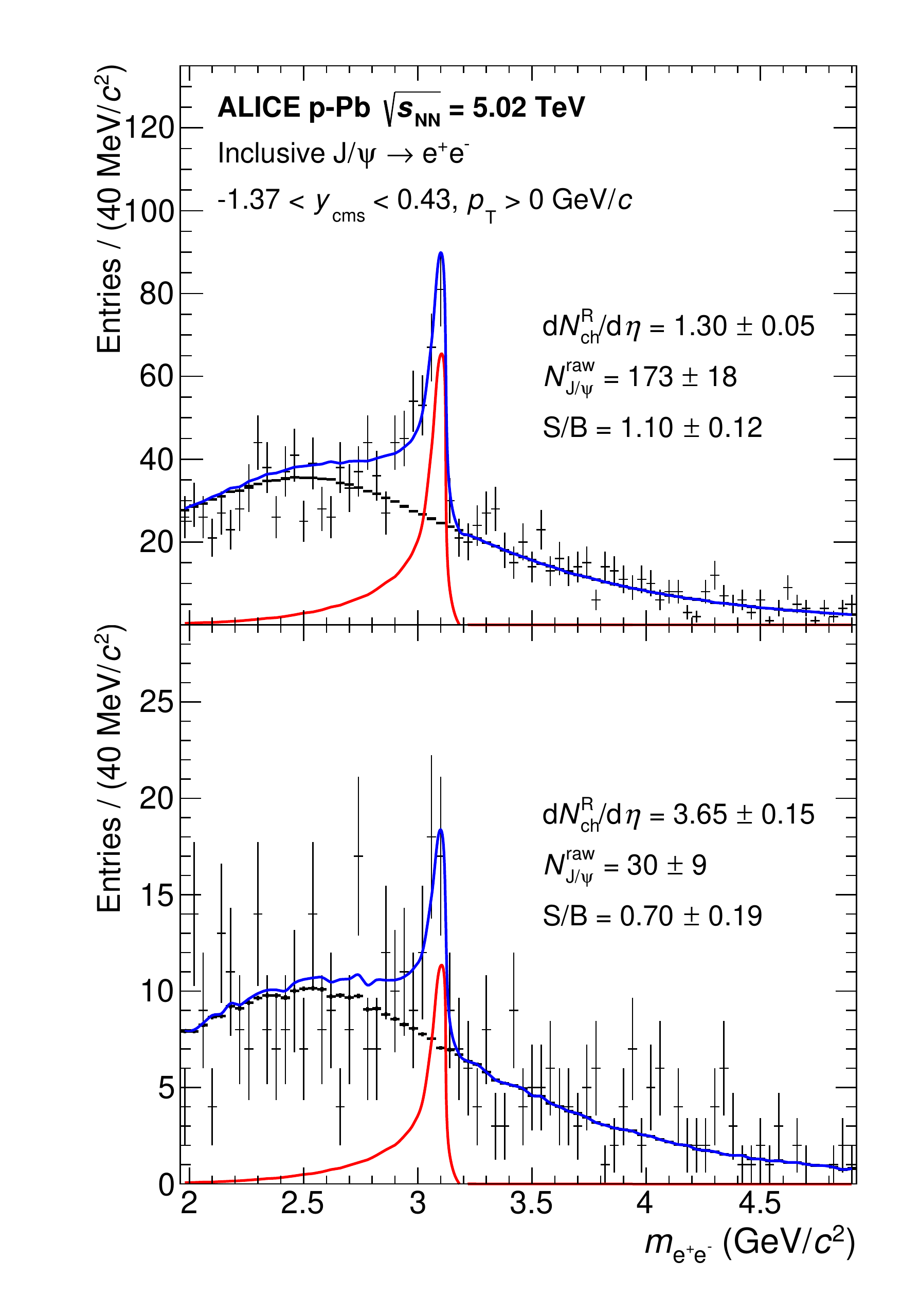}
\par}
\caption{\label{fig:minv}  Opposite-sign invariant mass distributions of selected muon (left panel, for the forward rapidity) and electron (right panel) pairs, for selected multiplicity bins. In the left panel, the distributions are corrected for \acef. The curves show the fit functions for signal, background and combined signal with background (see text for details). In the right panel, the background is evaluated with the event-mixing technique, and the overlaid signal is obtained from Monte Carlo (see text for details).}
\end{figure}

When building the invariant mass distributions, each dimuon pair (of a given \pt ~and \y) is corrected by the detector acceptance times efficiency factor 1/(\acef(\pt,\y)). The  \acef(\pt,\y) map is obtained from a particle-gun Monte Carlo (MC) simulation based on GEANT~3~\cite{GEANT3} and simulating the detector response as in \cite{Abelev:2014aa}. Since  the \acef-factor does not depend on the multiplicity for the event multiplicities relevant for the \ppb analyses, the simulated events only contain a dimuon pair at the generator level.  The simulations assume an unpolarised \jpsi production. The same reconstruction procedure and selection cuts are applied to MC events and to real data.
The extraction of the \jpsi signal in the dimuon channel is performed via a fit to the \acef-corrected opposite-sign (OS) dimuon invariant mass distributions obtained for \pt ~$<$ 15 GeV/$c$. The fitting procedure is similar to that used in a previous \jpsi analysis in \ppb collisions \cite{Abelev:2014aa}. The distributions are fitted using a superposition of \jpsi and \psip~signals and a background shape. 
The resonances are parameterized using a Crystal Ball function with asymmetric tails while for the background a gaussian with its width linearly varying with mass is used. In the present analysis, the parameters of the non-gaussian tails of the resonance shape are determined from fits of the MC \jpsi signal, and fixed in the data fitting procedure. Examples of fits of the \acef-corrected dimuon invariant mass distributions for two selected bins, low and high multiplicities, are given in the left panel of Fig.~\ref{fig:minv}.

In the dielectron decay channel, electrons and positrons are reconstructed in the central barrel detectors by requiring a minimum of 70 out of maximally 159 track points in the TPC and a maximum value of 4 for the track fit $\chi^2$ over the number of track points. Furthermore, only tracks with at least two associated hits in the ITS, one of them in the innermost layer, are accepted.  This selection reduces the amount of electrons and positrons from photon conversions in the material of the detector beyond the first ITS layer. In addition a veto cut on topologically identified tracks from photon conversions is applied. The electron identification is achieved by the measurement of the energy deposition of the track in the TPC, which is required to be compatible with that expected for electrons within 3 standard deviations.  Tracks with specific energy loss being consistent with that of the pion or proton hypothesis within 3.5 standard deviations are rejected. These selection criteria are identical to those used in~\cite{Adam:2015jsa}. Electrons and positrons are selected in the pseudorapidity range $|\eta|< $ 0.9 and in the transverse momentum range \pt ~$>$ 1 GeV/$c$. 

The background in the OS invariant mass distribution is estimated with dielectron pairs formed with tracks from different events (mixed-event background). The background shape is normalised such that its integral over ranges of the invariant mass in the sidebands of the \jpsi~mass peak equals the number of measured OS dielectron pairs in the same ranges (typical ranges used are $[3.2,3.7]$ GeV/$c^2$ and $[2.0,2.5]$ GeV/$c^2$). The signal itself is extracted by counting the entries in the background-subtracted invariant mass distribution (the standard range used is $[2.92,3.16]$ GeV/$c^2$). Due to bremsstrahlung of the electron and positron in the detector material and radiative corrections of the decay vertex, the J/$\psi$ signal shape has a tail towards lower invariant masses.
 The standard range for the signal extraction contains, according to MC simulations, about 69\% of the J/$\psi$ signal. The number of reconstructed J/$\psi$ mesons and its statistical uncertainty are derived from the mean obtained when varying the counting window for the signal extraction and the invariant mass ranges used for the normalisation of the background. The variations that are taken into account are the same as in~\cite{Adam:2015jsa}. 
Examples of the dielectron invariant mass distributions in data, for two selected analysis bins at low and high multiplicities, are given in the right panel of Fig.~\ref{fig:minv}.

The correction for the acceptance and efficiency of the raw yields is based on simulated \ppb collisions with the HIJING event generator~\cite{Wang:1991hta} with an injected J/$\psi$ signal.  The dielectron decay is simulated with the EVTGEN package~\cite{Lange:2001uf} using PHOTOS~\cite{BARBERIO1991115,BARBERIO1994291} to describe the final state radiation. The production is assumed to be unpolarised as in the muon decay channel analysis. The propagation of the simulated particles is done by GEANT~3~\cite{GEANT3} and a full simulation of the detector response is performed. The same reconstruction procedure and selection cuts are applied to MC events and to real data.

The inclusive \jpsi yield per event is obtained in each multiplicity bin as
$N_{\mathrm{J/\psi}} = {N_{\mathrm{J/\psi}}^{\mathrm{corr}}}/{N_{\mathrm{MB}}}$,
where $N_{\mathrm{J/\psi}}^{\mathrm{corr}}$ is the number of reconstructed \jpsi mesons corrected for the acceptance times efficiency factor. In the dimuon decay channel analysis, the number of MB events equivalent to the analysed dimuon sample ($N_{\mathrm{MB}}$) in each multiplicity bin is obtained from the number of dimuon triggers ($N_{\mathrm{DIMU}}$), through the normalisation factor of dimuon-triggered to MB-triggered events $F_{2\mu / \mathrm{MB}}$, as $N_{\mathrm{MB}} = F_{2\mu / \mathrm{MB}} \cdot N_{\mathrm{DIMU}}$. This factor is computed using two different methods, as discussed in \cite{Adam:2015jsa}. The \jpsi %integrated 
cross section values for minimum-bias events obtained in the dimuon channel at forward and backward rapidities, and in the dielectron channel at mid-rapidity are compatible with those presented in \cite{Abelev:2014aa} and \cite{Adam:2015ac}, respectively.
The results presented here are provided relative to the yield in NSD events, $\langle {\rm d}N_{J/\psi}/{\rm d}y\rangle$.
%This allows a comparison with other collision systems at different energies. 
The event-averaged yield is normalised to the NSD event class; the normalisation uncertainty is 3.1\% \cite{Abelev:2013ab}. 

 \begin{figure}[htb]
  {\centering 
\includegraphics[width=0.5\textwidth]{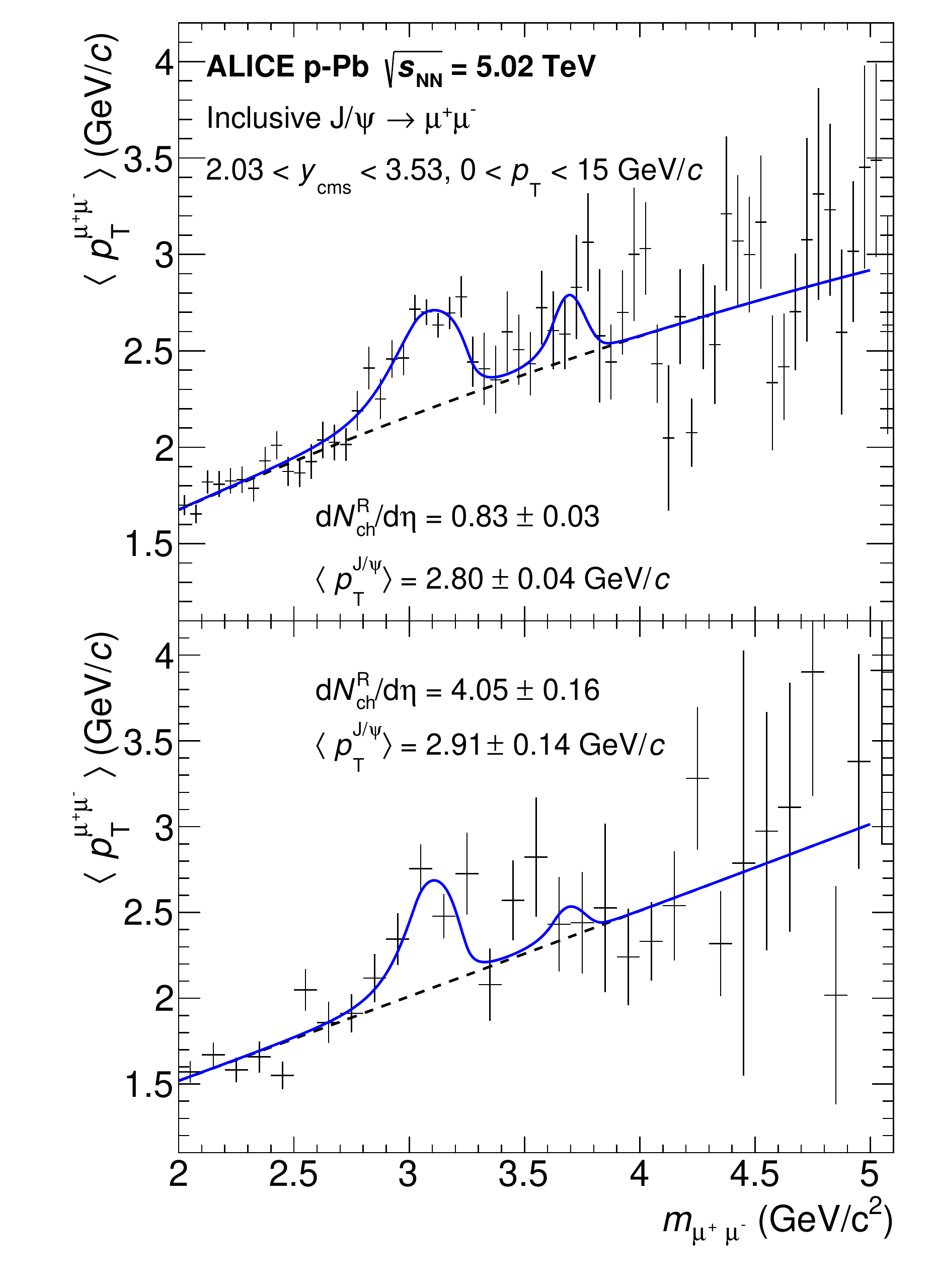}
\par}
\caption{\label{fig:mpt} Average transverse momentum of opposite-sign muon pairs as a function of the invariant mass at forward rapidity, for two multiplicity bins. The curves are fits of the background and combined signal and background (see text).}
\end{figure}

In previous analyses, e.g. \cite{Adam:2015ac}, the \jpsi yield was extracted in \pt ~bins, and the resulting distribution was fitted to extract the \mpt value. The present analysis aims at studying effects that may arise at high charged-particle multiplicities, where the usual method is no longer suitable due to statistical limitations. The method presented here does not require to sample the data in \pt ~bins, hence allowing the analysis in finer multiplicity bins. 
The extraction of the average transverse momentum of \jpsi mesons is done via a fit to the dimuon mean transverse momentum as a function of the invariant mass, $\average{p_{\rm{T}}^{\mu^{+}\mu^{-}}} (m_{\mu^{+}\mu^{-}})$.  A correction for the acceptance times efficiency has to be applied when building these distributions. Hence, the contribution of each dimuon pair with a certain \pt ~and $y$ in a given invariant mass bin is weighted with the two-dimensional \acef(\pt,\y). 
In order to extract the \jpsi \mpt, the \acef-corrected $\average{\pt^{\mu^{+}\mu^{-}}} (m_{\mu^{+} \mu^{-}})$ distributions, which are shown in Fig.~\ref{fig:mpt}, are fitted using the following functional shape:

\begin{eqnarray} \label{mptfitfcn}
\average{\pt^{\mu^{+} \mu^{-}}} (m_{\mu^{+} \mu^{-}}) &=& \alpha^{ J/\psi }(m_{\mu^{+} \mu^{-}}) \times \average{\pt^{J/\psi}}
\nonumber  \\
&+& \alpha^{ \psi(2S) }(m_{\mu^{+} \mu^{-}}) \times \average{\pt^{\psi(2S)}}
\nonumber  \\
&+& \left( 1 - \alpha^{J/\psi}(m_{\mu^{+} \mu^{-}}) -  \alpha^{\psi(2S)}(m_{\mu^{+} \mu^{-}}) \right) \times \average{\pt^{\mathrm{bkg}}}
\end{eqnarray}

where $\alpha(m_{\mu^{+} \mu^{-}}) = S(m_{\mu^{+} \mu^{-}}) / (S(m_{\mu^{+} \mu^{-}}) +B(m_{\mu^{+} \mu^{-}}))$; the signal ($S$) and background ($B$) dependence on the dimuon invariant mass is extracted from the corrected invariant mass spectrum fits mentioned above. The \jpsi and \psip ~average transverse momenta, $\average{\pt^{J/\psi}}$ and $\average{\pt^{\psi(2S)}}$, respectively, are fit parameters assumed to be independent of the invariant mass, while the background one, $\average{\pt^{\mathrm{bkg}}}$, is parameterized with a second order polynomial function. Note that, as for the yield extraction, the quantity $\average{\pt^{\psi(2S)}}$ is not a measurement of the $\psi(2\mathrm{S})$ mean transverse momentum, since the \acef ~is obtained only from \jpsi signals in the simulation. 
The \mpt ~results presented here for bins in multiplicity are relative to the value obtained for inclusive events, $\mpt_{\mathrm{MB}}$ \cite{Adam:2015ac}.

\section{Systematic uncertainties}
\label{sec:syst}
%!TEX root = ../JpsiVsChMult.tex

The systematic uncertainty of the overall average charged-particle pseudorapidity density was estimated to be 3.8\% \cite{Abelev:2013ab}. This includes effects related to the uncertainties in the simulations, detector acceptance and event selection efficiency, and it is dominated by the normalisation to the NSD event class. 
Possible correlation between the average multiplicity and that evaluated in a given bin would lead to a partial cancellation of certain sources of uncertainty when computing the relative multiplicity. As a conservative estimate, the uncertainty on the relative multiplicity is considered to be equal to the uncertainty on the overall charged-particle pseudorapidity density.

The influence of variations of the $\eta$ distribution in the calculation of the $\beta$ correction factors, is estimated from the difference between the average number of tracklets obtained in the data taken with the two different beam configurations. 
The corresponding uncertainty on the multiplicity determination amounts to 1\%. The uncertainties arising from the fit procedure of the \ntrcorr - \nch ~correlation in simulated events, used to obtain the correction factors, are also included. This uncertainty ranges between 0.2\% (at high multiplicity) and 2\% (at low multiplicity). The event selection related to the vertex quality has a 1\% effect on the average multiplicity in the lowest multiplicity bin and a negligible effect for the other bins. Due to the uncertainty on the determination of the multiplicity of the individual events, there could be a migration of events among the multiplicity bins (bin-flow). This bin-flow effect is determined by running the analysis several times with different seeds for the random factor of the multiplicity correction (bin-flow test). The bin-flow uncertainties are obtained from the dispersion of the average multiplicity values in the bin-flow tests for each multiplicity bin. Finally, the effect of pile-up is studied using a toy model that reproduces the main features of the multiplicity determination, and takes into account the mis-identification of multiple collisions in the same event. The contributions of bin-flow and pile-up to the measured multiplicities are found to be negligible for all the data sets (taken at different interaction rates). 
The bin-dependent uncertainty is added in quadrature to the 3.8\% uncertainty of  $\average{\dndeta}$, resulting in a systematic uncertainty of the relative charged-particle multiplicity of 4 $-$ 4.5\% depending on the multiplicity bin.

The yields reported here are provided relative to the event-average yield and the uncertainties are estimated for this ratio. The systematic uncertainties related to trigger, tracking and matching efficiency are correlated between the multiplicity-differential and the integrated determinations. They cancel out to a large extent. 

In the dimuon analysis, a combined systematic uncertainty which includes the \acef ~variations due to the uncertainty of the \jpsi \pt ~and rapidity input distributions used in the simulation and multiplicity bin-flow effects is derived. Due to the multiplicity bin-flow, and the fact that the invariant mass and $\average{p_{\rm{T}}^{\mu^{+}\mu^{-}}} (m_{\mu^{+}\mu^{-}})$ spectra are weighted by \acef, these uncertainties can not be computed separately. The combined uncertainty is obtained from the \rms ~~of the relative yield values obtained running the analysis several times with different seeds for the random factor of the multiplicity correction. In addition the systematic uncertainty for the signal extraction is estimated as the \rms ~~of the results obtained using different fitting assumptions for a given bin-flow test. 
The fit procedure is varied by adopting a pseudo-gaussian function for the signal, a polynomial times an exponential function for the background and by using two additional fitting ranges. The uncertainties due to the determination of the parameters of the signal tails is estimated by using several sets of parameters from different MC simulations. 
The uncertainty related to the computation method of the relative $F_{2\mu / MB}$ is estimated considering the difference between the two available methods to measure the factor in multiplicity bins \cite{Adam:2015jsa}. 
The effect of the vertex quality selection is estimated from the difference of the obtained yields with and without this selection. Finally, in order to determine the pile-up effect on the measured yield in each multiplicity bin, the pile-up toy model is extended by including the production of \jpsi using as input the measured yields as a function of multiplicity. The difference between the measured and toy MC yields is taken as systematic uncertainty. All these effects are uncorrelated within a given multiplicity bin, hence they are added quadratically to obtain the systematic uncertainty of the relative yield in a multiplicity bin. Also, these systematic uncertainties are considered as uncorrelated between the different rapidity intervals. A summary of the maximum and minimum relative yield systematic uncertainties is shown in Tab.~\ref{tab:jpsisystComb}. In addition, the 3.1\% uncertainty of the event-average yield normalisation to NSD, is reported separately. 

The systematic uncertainties are computed also for the absolute yields in multiplicity bins at forward and backward rapidities. The absolute yields are used to compute the ratio of the nuclear modification factors at forward and backward rapidities. The values of the uncertainties on the absolute yields are shown in parentheses in Tab.~\ref{tab:jpsisystComb}, when they are different from the ones obtained for the relative yield. In addition, for the absolute yield measurement, the muon tracking, trigger and matching efficiency uncertainties need to be taken into account \cite{Abelev:2014aa}. They amount to 4\% (6\%), 3\% (3.4\%) and 1\% (1\%) at forward (backward) rapidities.

\begin{table} [h!]
\centering
\begin{tabular}{cccc}
\toprule
Source & 2.03 $< y_{\mathrm{cms}} <$ 3.53 & $-$4.46 $< y_{\mathrm{cms}} < -$2.96 & $-$1.37 $< y_{\mathrm{cms}} <$ 0.43   \\
\midrule
Sig. Extr. & 0.6 $-$ 1.9\%  & 0.5 $-$ 1.8\% &  3.2 $-$ 8.4\% \\
$F_{2\mu / MB}$ method & 0.3 $-$ 3.9\% & 0.3 $-$ 3.9\% & not applicable \\
\acef/bin-flow & 1 $-$ 5.9\% & 1.4 $-$ 3.9\% &   3.1 $-$ 9.9\%  \\
Pile-up & 1 $-$ 4\% & 1 $-$ 1.5\% & negligible\\
Signal tails & 0.5\% (2\%) & 0.5\% (2\%) & not applicable\\
Vertex quality sel. & 0.3\%-0.6\%$^{*}$ (0.9\%$^{*}$) & 0.3\%-0.6\%$^{*}$ (0.9\%$^{*}$)& negligible \\
%\addlinespace[\spacebeforetotal]
\hline %\hline
Total & 2.1 $-$ 8.3\% (3.0 $-$ 8.6\%) & 2.1 $-$ 6.0\% (2.9 $-$ 6.4\%)  & 4.5 $-$ 13\% \\
\bottomrule
\end{tabular}
\caption{\label{tab:jpsisystComb} The relative systematic uncertainties  of the relative \jpsi ~yield measurement in the three rapidity ranges. The values in parentheses correspond to the absolute yield measurement when different from the relative ones. The ranges represent the minimum and maximum values of the uncertainties over the multiplicity bins. For the vertex quality selection, the uncertainties marked with $^{*}$ refer only to the lowest-multiplicity bin; for all other bins the value is 0.3\%. The trigger, tracking and matching efficiency uncertainties are not listed in the table.}
\end{table}

For the dielectron decay channel, the signal extraction uncertainty is derived based on the \rms  ~~value of the  different signal yield ratios  obtained for the variations of the background  and the signal integration window as in \cite{Adam:2015jsa}.  The uncertainty is largest for the highest multiplicity bins. Since the \pt ~distribution of \jpsi may depend on multiplicity, the unmeasured \pt ~spectrum leads to a multiplicity-dependent uncertainty, determined as in \cite{Adam:2015jsa}. As explained in section~\ref{sec:data}, the pile-up contamination is very low and the induced uncertainty is negligible for all the multiplicity intervals.
The uncertainty related to bin-flow is estimated with the same method as in the dimuon analysis.
The total systematic uncertainty varies as a function of multiplicity between 4.5\% and 13\%, see Tab.~\ref{tab:jpsisystComb}.

For the relative \jpsi \mpt, the effects of the uncertainty on the determination of \acef, the \mpt ~extraction procedure and bin-flow are computed together following the same procedure as for the relative yield. The \mpt ~extraction uncertainty is obtained from the dispersion of the results using different fit combinations, including variations of the invariant mass signal and background parameterisations, fitting range and the use of a second order polynomial times an exponential function for the \mpt of background dimuons. The effect of considering the \jpsi \mpt ~as independent of the invariant mass in the $\langle p_{\rm{T}}^{\mu^{+}\mu^{-}} \rangle$ fits is found to be negligible. The impact of fixing the signal and background parameters during the fitting procedure is observed to be negligible as well. The events removed by the vertex quality selection do not have reconstructed \jpsi and therefore the \mpt ~remains unmodified. Finally, using a pile-up toy model, it is shown that the pile-up has no effect on the \mpt ~measurement, except for the two bins corresponding to the largest multiplicities. All these effects are considered as uncorrelated in a given multiplicity bin and hence their respective uncertainties are added quadratically to obtain the relative \mpt ~systematic uncertainty in each multiplicity bin. These systematic uncertainties are considered as uncorrelated between the different rapidity intervals. The results of the uncertainties entering in the relative \mpt ~measurement are reported in Tab.~\ref{tab:jpsimptsystComb}. 

\begin{table} [h!]
\centering
\begin{tabular}{ccc}
\toprule
Source & 2.03 $< y_{\mathrm{cms}} <$ 3.53 & $-$4.46 $< y_{\mathrm{cms}} < -$2.96\\
\midrule
Sig. extr. & 0.2 $-$ 1.2\% & 0.2 $-$ 0.5\% \\
\acef/bin-flow & 0.5 $-$ 2.0\% & 0.7 $-$ 2.7\%\\
Pile-up & 0.2 $-$ 0.7\%$^{*}$ & 0.2 $-$ 0.8\%$^{*}$  \\
%\addlinespace[\spacebeforetotal]
\hline %\hline
Total & 0.6 $-$ 2.4\% &  0.7 $-$ 2.9\% \\
\bottomrule
\end{tabular}
\caption{\label{tab:jpsimptsystComb} The systematic uncertainties for the relative \mpt ~measurement at forward and backward rapidities. The values represent the minimum and maximum values of the uncertainties over the multiplicity bins. The uncertainties marked with $^{*}$ only refer to the two highest-multiplicity bins.}
\end{table}

\section{Results and discussion}
\label{sec:res}
%!TEX root = ../JpsiVsChMult.tex

 \begin{figure}[htb]
  {\centering 
\resizebox*{.65\columnwidth}{!}{\includegraphics{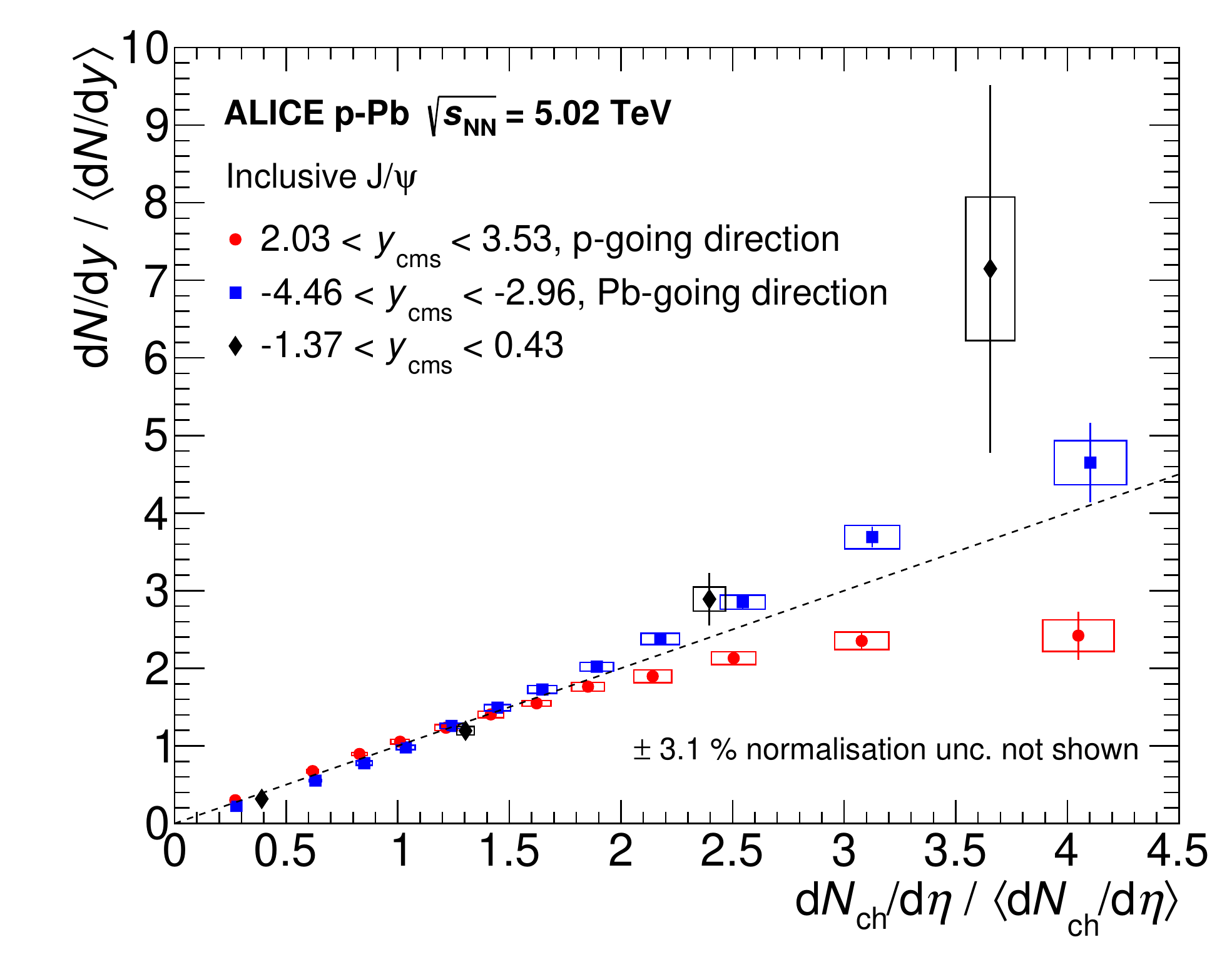}} 
\par}
\caption{\label{fig:jpsiyield} Relative yield of inclusive \jpsi mesons, measured in three rapidity regions, as a function of relative charged-particle pseudorapidity, measured at mid-rapidity.
The error bars show the statistical uncertainties, and the boxes the systematic ones.
The dashed line is the first diagonal, plotted to guide the eye.
}
\end{figure}

The dependence of the relative \jpsi yield on the relative charged-particle pseudorapidity density for three \jpsi rapidity ranges is presented in Fig.~\ref{fig:jpsiyield}. An increase of the relative yield with charged-particle multiplicity is observed for all rapidity domains, with a similar behaviour at low multiplicities. At multiplicities beyond 1.5$-$2 times the event-average multiplicity, two different trends are observed. The relative yields at mid-rapidity and backward rapidity keep growing with the relative multiplicity in \ppb collisions similarly to the observation in \pp ~collisions at 7 TeV \cite{Abelev:2012aa}. At forward rapidity the trend is different. In this rapidity window a saturation of the relative yield sets in for high multiplicities. In lack of theoretical model calculations, it is unclear at the moment what is the cause of this observation.
We recall that the explored Bjorken $x$ ranges in the forward rapidity region are in the domain of shadowing/saturation, and that a variety of models \cite{Vogt:2013aa,Arleo:2013aa,Ferreiro:2014bia,Ma:2015sia,Ducloue:2015gfa} are fairly successful in describing the recent centrality-integrated and differential measurements of ALICE \cite{Adam:2015ac,Adam:2015jsa}, which correspond in terms of our relative multiplicities to $\dndeta/\average{\dndeta}\simeq2.5$ at most.

\begin{figure}[htb]
  {\centering 
\resizebox*{.65\columnwidth}{!}{\includegraphics{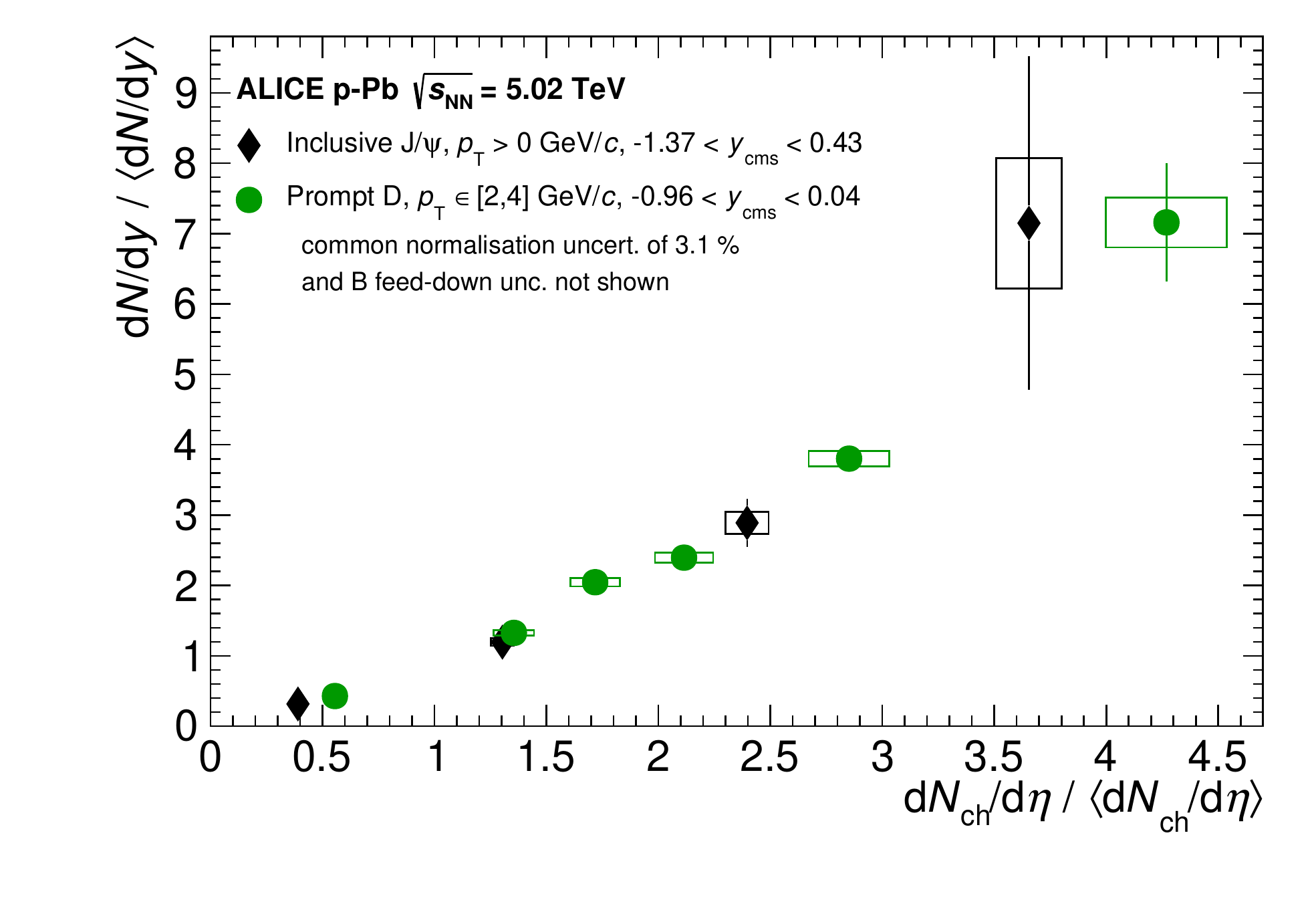}}
\par}
\caption{\label{fig:jpsi2d} Relative yield of inclusive \jpsi mesons as a function of relative charged-particle pseudorapidity density, measured at mid-rapidity, in comparison to D mesons (average of D$^0$, D$^+$, and D$^{*+}$ species), for the \pt ~interval 2-4 \gevc ~\cite{Adam:2016mkz}.
The error bars show the statistical uncertainties, and the boxes the systematic ones (additional systematic uncertainties due to the b feed-down contributions and the event normalisation are not shown for the D mesons).
}
\end{figure}

In Fig.~\ref{fig:jpsi2d} the \jpsi measurement at mid-rapidity is compared to that for prompt D mesons (average of D$^0$, D$^+$, and D$^{*+}$ species) for the \pt~ range 2$-$4 \gevc~ \cite{Adam:2016mkz}. Similar trends are seen for the \jpsi and D mesons, as observed earlier in pp collisions \cite{Adam:2015ota}. 

The nuclear modification factor for \jpsi production in \ppb collisions (\rpb) as a function of centrality was presented in \cite{Adam:2015jsa}. The relationship between geometry-related quantities, that quantify the centrality of the collision, and experimental observables in \ppb collisions may be subject to a selection bias \cite{Adam:2015aa} which needs care in interpretation. 
By performing the ratio of the nuclear modification factors at forward and backward rapidities as a function of multiplicity, the dependence on geometry-related quantities is eluded. The forward-to-backward nuclear modification factor ratio is defined as:

\begin{eqnarray} \label{rfb}
R_{\mathrm{FB}} &=& \frac{\rpb(2.03 < \ycms < 3.53)}{\rpb(-4.46 < \ycms < -2.96)} \\ \nonumber
 &=& \frac{Y^{J/\psi}_{\mathrm{pPb}} (2.03 < \ycms < 3.53)}{Y^{J/\psi}_{\mathrm{pPb}} (-4.46 < \ycms < -2.96)} \times
    \frac{\mathrm{d}\sigma^{J/\psi}_{\mathrm{pp}}/\mathrm{d}y (-4.46 < \ycms <
    -2.96)}{\mathrm{d}\sigma^{J/\psi}_{\mathrm{pp}}/\mathrm{d}y (2.03 < \ycms < 3.53) }
\end{eqnarray}

Since the average charged-particle multiplicities and their uncertainties are consistent with each other for the two sets of data, the values of $R_{\mathrm{FB}}$ are shown versus the average value of the two in each multiplicity bin. Note that, differently than for the case of the nuclear modification factor measurement in \cite{Abelev:2014aa}, for the present measurement the rapidity ranges are not symmetric with respect to \ycms ~= 0 to take advantage of all the signal yield, allowing the study up to high multiplicities. 
The values of the reference \pp ~cross section were obtained by means of an interpolation procedure using measurements at center-of-mass energies of 2.76 and 7 TeV \cite{ALICELHCbRefpp}. The resulting backward-to-forward ratio of \jpsi production cross sections in \pp ~collisions is $ 0.691 \pm 0.048$, leading to a global uncertainty on the $R_{\mathrm{FB}}$ measurement of 6.9\%.

For the $R_{\mathrm{FB}}$ ratio, the systematic uncertainties of the absolute yields in \ppb collisions (Tab.~\ref{tab:jpsisystComb}) are considered as uncorrelated between forward and backward rapidities, and therefore added in quadrature. The uncorrelated systematic uncertainties of the production cross sections in \pp ~collisions are the same as a function of multiplicity, so they are added in quadrature to the global uncertainty (quadratic sum of muon tracking, trigger and matching efficiency uncertainties) of the \ppb data, resulting in a total relative uncertainty of 11\%.

 \begin{figure}[htb]
  {\centering 
\resizebox*{.67\columnwidth}{!}{\includegraphics{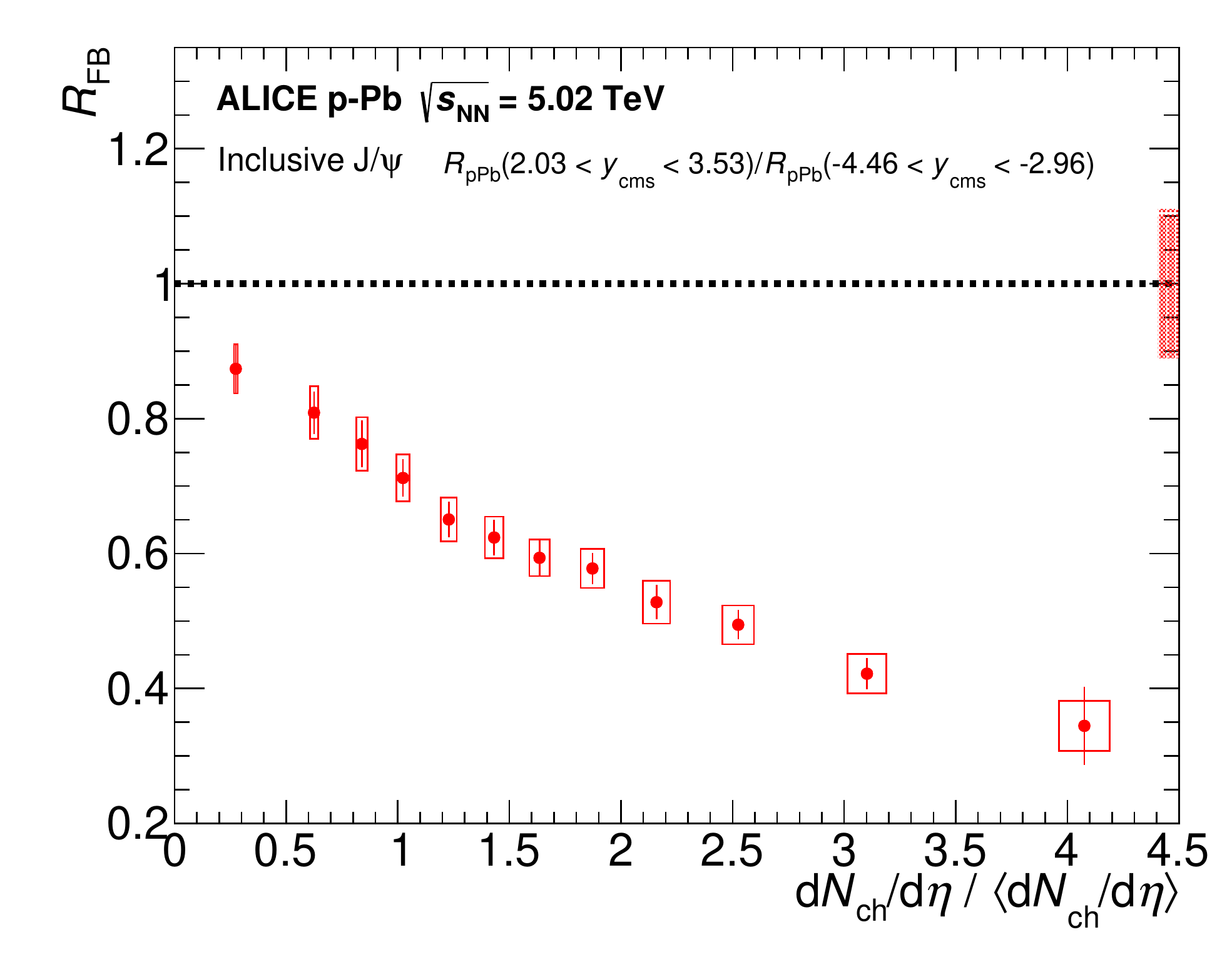}} 
\par}
\caption{\label{fig:RFBvsMult}  $R_{\mathrm{FB}}$ of inclusive \jpsi in p-Pb collisions at $\sqrt{s_{\mathrm{NN}}} = $ 5.02 TeV as a function of relative charged-particle pseudorapidity density, measured at mid-rapidity. The red box around unity represents the global uncertainty.
The error bars show the statistical uncertainties, and the boxes the systematic ones.
}
\end{figure}

The $R_{\mathrm{FB}}$ ratio is shown as a function of the relative charged-particle pseudorapidity density in Fig.~\ref{fig:RFBvsMult}. In multiplicity-inclusive collisions 
for symmetric $y$ ranges at forward and backward rapidities \cite{Abelev:2014aa}, $R_{\mathrm{FB}}$ is smaller than unity and described by theoretical models. The present measurement shows that the suppression of \jpsi production at forward rapidity with respect to backward rapidity increases significantly with charged-particle multiplicity, since $R_{\mathrm{FB}}$ reaches values as low as 0.34 $\pm$ 0.06 (stat.) $\pm$ 0.05 (syst.). 
A forward-backward asymmetry can be noticed for inclusive charged-particle production studied in \cite{Adam:2015aa}. 
Even though the range of relative charged-particle multiplicities probed in that measurement is not as large as in the present measurement of \jpsi production, the apparent similarity of the trend seen in Fig.~\ref{fig:RFBvsMult} to soft particle production is intriguing.

\begin{figure}[htb]
  {\centering 
\resizebox*{.68\columnwidth}{!}{\includegraphics{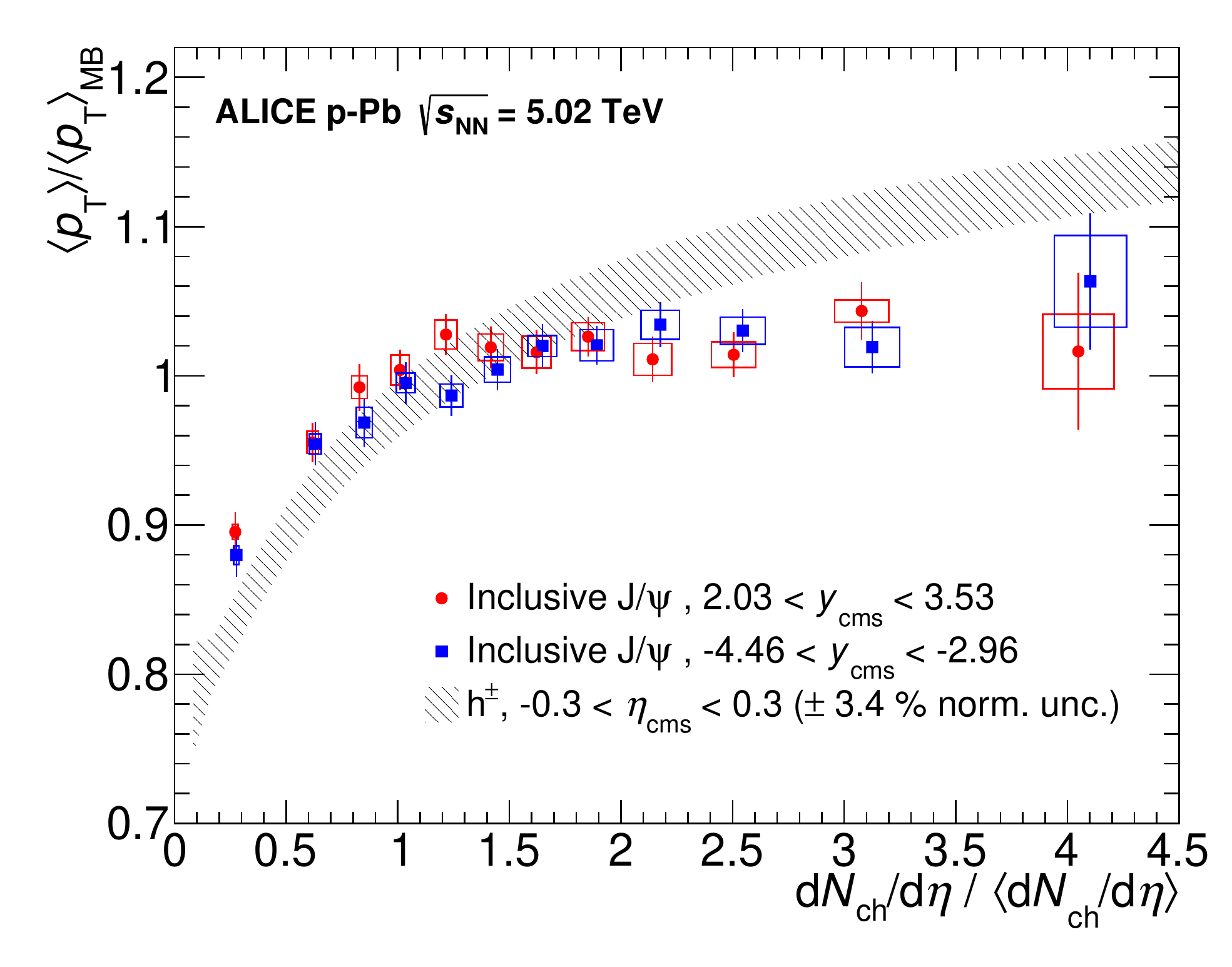}}
\par}
\caption{\label{fig:jpsimpt} Relative \mpt ~of \jpsi mesons for backward and forward rapidity as a function of the relative charged-particle pseudorapidity density, measured at mid-rapidity. The bars show the statistical uncertainties, and the boxes the systematic ones. The data for charged particles (h$^\pm$) \cite{Abelev:2013aa} are included for comparison. The latter are for $|\etacms|<0.3$ and with \pt ~in the range 0.15 to 10 \gevc ~and have an additional normalisation uncertainty of 3.4\%.
}
\end{figure}

In Fig.~\ref{fig:jpsimpt} the relative \mpt ~of \jpsi mesons at backward and forward rapidity is shown as a function of the relative charged-particle pseudorapidity density. The results are similar at forward and backward rapidities. An increase of the relative \mpt ~with multiplicity at low charged-particle multiplicity is observed, but for multiplicities beyond 1.5 times the average multiplicity it saturates.  For backward rapidity, the simultaneous increase of the yield and the saturation of the relative \mpt could be an indication of \jpsi production from an incoherent superposition of parton-parton interactions, as suggested by data on correlations of jet-like yields per trigger particle \cite{Abelev:2015aa}.

The \pt ~broadening observed in the analysis of \jpsi production in \ppb collisions as a function of centrality \cite{Adam:2015jsa} is well described by initial and final-state multiple scattering of partons within the nuclear medium \cite{Kang:2013aa}. The comparison of data to model calculations, performed in \cite{Adam:2015jsa}, corresponds in terms of relative multiplicities to a range up to roughly $\dndeta/\average{\dndeta}=2.5$. It remains to be seen whether such models can explain the saturation observed in the relative \mpt ~of the \jpsi mesons for events with higher multiplicities.

It is interesting to contrast the observed saturation of \mpt~for \jpsi mesons with the monotonic increase of \mpt ~for charged hadrons (dominated by pion production) ~\cite{Abelev:2013bla} with the multiplicity measured at mid-rapidity also shown in Fig.~\ref{fig:jpsimpt}. Note that this measurement is for particles in $|\etacms|<0.3$ and with \pt ~in the range 0.15 to 10 \gevc, and it is relative to events with at least one particle in this kinematic range (for which $\average{\nch}=11.9 \pm 0.5$ and $\average{\pt}=0.696 \pm 0.024$ \gevc~\cite{Abelev:2013bla}).
Although the different kinematic regions may play a role and care is needed in the interpretation, it is apparent that the two observables, characterised by rather different production mechanisms (and momentum-transfer) exhibit different patterns in the multiplicity dependence of the average transverse momentum.

\section{Conclusions}
\label{sec:conc}
%!TEX root = ../JpsiVsChMult.tex

Measurements of the relative \jpsi yield and average transverse momentum as a function of the relative charged-particle pseudorapidity density in \ppb collisions at the LHC at $\snn = 5.02$ TeV have been presented in this letter.
The measurements were performed with ALICE in three ranges of rapidity. The charged-particle multiplicity was measured at mid-rapidity; multiplicities up to 4 times the value of NSD events were reached, corresponding to rare events of less than 1\% of the total hadronic interaction cross section.
An increase of the relative \jpsi yield with the relative multiplicity is observed, with a trend towards saturation at high multiplicity for the forward rapidity (proton-going direction). 
For the \jpsi data at mid-rapidity, a comparison to corresponding measurements of D-meson yields is performed, revealing similar patterns for the two meson species.
At forward and backward rapidities, the relative average transverse momenta exhibit a saturation above moderate values of relative multiplicity.

The present data are expected to constitute a stringent test for theoretical models of \jpsi production in \mbox{\ppb} collisions and help to understand the effects associated with the production of a deconfined medium in \pb collisions. 
%The measurements reported here may help to clarify whether such a medium is also produced in high-multiplicity \ppb collisions.

%
%

%%%%% acknowledgements
\newenvironment{acknowledgement}{\relax}{\relax}
\begin{acknowledgement}
\section*{Acknowledgements}
% Version: 2017-01-10

The ALICE Collaboration would like to thank all its engineers and technicians for their invaluable contributions to the construction of the experiment and the CERN accelerator teams for the outstanding performance of the LHC complex.
The ALICE Collaboration gratefully acknowledges the resources and support provided by all Grid centres and the Worldwide LHC Computing Grid (WLCG) collaboration.
The ALICE Collaboration acknowledges the following funding agencies for their support in building and running the ALICE detector:
A. I. Alikhanyan National Science Laboratory (Yerevan Physics Institute) Foundation (ANSL), State Committee of Science and World Federation of Scientists (WFS), Armenia;
Austrian Academy of Sciences and Nationalstiftung f\"{u}r Forschung, Technologie und Entwicklung, Austria;
Ministry of Communications and High Technologies, National Nuclear Research Center, Azerbaijan;
Conselho Nacional de Desenvolvimento Cient\'{\i}fico e Tecnol\'{o}gico (CNPq), Universidade Federal do Rio Grande do Sul (UFRGS), Financiadora de Estudos e Projetos (Finep) and Funda\c{c}\~{a}o de Amparo \`{a} Pesquisa do Estado de S\~{a}o Paulo (FAPESP), Brazil;
Ministry of Science \& Technology of China (MSTC), National Natural Science Foundation of China (NSFC) and Ministry of Education of China (MOEC) , China;
Ministry of Science, Education and Sport and Croatian Science Foundation, Croatia;
Ministry of Education, Youth and Sports of the Czech Republic, Czech Republic;
The Danish Council for Independent Research | Natural Sciences, the Carlsberg Foundation and Danish National Research Foundation (DNRF), Denmark;
Helsinki Institute of Physics (HIP), Finland;
Commissariat \`{a} l'Energie Atomique (CEA) and Institut National de Physique Nucl\'{e}aire et de Physique des Particules (IN2P3) and Centre National de la Recherche Scientifique (CNRS), France;
Bundesministerium f\"{u}r Bildung, Wissenschaft, Forschung und Technologie (BMBF) and GSI Helmholtzzentrum f\"{u}r Schwerionenforschung GmbH, Germany;
Ministry of Education, Research and Religious Affairs, Greece;
National Research, Development and Innovation Office, Hungary;
Department of Atomic Energy Government of India (DAE) and Council of Scientific and Industrial Research (CSIR), New Delhi, India;
Indonesian Institute of Science, Indonesia;
Centro Fermi - Museo Storico della Fisica e Centro Studi e Ricerche Enrico Fermi and Istituto Nazionale di Fisica Nucleare (INFN), Italy;
Institute for Innovative Science and Technology , Nagasaki Institute of Applied Science (IIST), Japan Society for the Promotion of Science (JSPS) KAKENHI and Japanese Ministry of Education, Culture, Sports, Science and Technology (MEXT), Japan;
Consejo Nacional de Ciencia (CONACYT) y Tecnolog\'{i}a, through Fondo de Cooperaci\'{o}n Internacional en Ciencia y Tecnolog\'{i}a (FONCICYT) and Direcci\'{o}n General de Asuntos del Personal Academico (DGAPA), Mexico;
Nederlandse Organisatie voor Wetenschappelijk Onderzoek (NWO), Netherlands;
The Research Council of Norway, Norway;
Commission on Science and Technology for Sustainable Development in the South (COMSATS), Pakistan;
Pontificia Universidad Cat\'{o}lica del Per\'{u}, Peru;
Ministry of Science and Higher Education and National Science Centre, Poland;
Korea Institute of Science and Technology Information and National Research Foundation of Korea (NRF), Republic of Korea;
Ministry of Education and Scientific Research, Institute of Atomic Physics and Romanian National Agency for Science, Technology and Innovation, Romania;
Joint Institute for Nuclear Research (JINR), Ministry of Education and Science of the Russian Federation and National Research Centre Kurchatov Institute, Russia;
Ministry of Education, Science, Research and Sport of the Slovak Republic, Slovakia;
National Research Foundation of South Africa, South Africa;
Centro de Aplicaciones Tecnol\'{o}gicas y Desarrollo Nuclear (CEADEN), Cubaenerg\'{\i}a, Cuba, Ministerio de Ciencia e Innovacion and Centro de Investigaciones Energ\'{e}ticas, Medioambientales y Tecnol\'{o}gicas (CIEMAT), Spain;
Swedish Research Council (VR) and Knut \& Alice Wallenberg Foundation (KAW), Sweden;
European Organization for Nuclear Research, Switzerland;
National Science and Technology Development Agency (NSDTA), Suranaree University of Technology (SUT) and Office of the Higher Education Commission under NRU project of Thailand, Thailand;
Turkish Atomic Energy Agency (TAEK), Turkey;
National Academy of  Sciences of Ukraine, Ukraine;
Science and Technology Facilities Council (STFC), United Kingdom;
National Science Foundation of the United States of America (NSF) and United States Department of Energy, Office of Nuclear Physics (DOE NP), United States of America.
    %%%%%%% done by webmaster team
\end{acknowledgement}

%%%%%%%% Bibliography (In case of using bibtex generate the bbl requested by arXiv)
\bibliographystyle{utphys}   % Remember we use title in the biblio
\bibliography{JpsiVsChMult}
%\input {bibliography.tex}  

%%%%%%%%% appendix with author list
\newpage
\appendix
\section{The ALICE Collaboration} \label{app:collab}

% Collaboration: CERN-LHC-ALICE
% Generation Date is 2017-Jan-10

% How to use:
%%%%%%%%% appendix with author list
%\appendix
%\section{The ALICE Collaboration}
%\label{app:collab}
%\input{authors-list.tex}  %%%%%%% get the latest version before submitting

\begingroup
\small
\begin{flushleft}
D.~Adamov\'{a}$^\textrm{\scriptsize 87}$,
M.M.~Aggarwal$^\textrm{\scriptsize 91}$,
G.~Aglieri Rinella$^\textrm{\scriptsize 34}$,
M.~Agnello$^\textrm{\scriptsize 30}$,
N.~Agrawal$^\textrm{\scriptsize 47}$,
Z.~Ahammed$^\textrm{\scriptsize 139}$,
N.~Ahmad$^\textrm{\scriptsize 17}$,
S.U.~Ahn$^\textrm{\scriptsize 69}$,
S.~Aiola$^\textrm{\scriptsize 143}$,
A.~Akindinov$^\textrm{\scriptsize 54}$
% ...
\renewcommand\labelenumi{\textsuperscript{\theenumi}~}

\section*{Affiliation notes}
\renewcommand\theenumi{\roman{enumi}}
\begin{Authlist}
\item \Adef{0}Deceased
\item \Adef{idp1795904}{Also at: Georgia State University, Atlanta, Georgia, United States}
\item \Adef{idp3231008}{Also at: Also at Department of Applied Physics, Aligarh Muslim University, Aligarh, India}
\item \Adef{idp4002544}{Also at: M.V. Lomonosov Moscow State University, D.V. Skobeltsyn Institute of Nuclear, Physics, Moscow, Russia}
\end{Authlist}

\section*{Collaboration Institutes}
\renewcommand\theenumi{\arabic{enumi}~}

$^{1}$A.I. Alikhanyan National Science Laboratory (Yerevan Physics Institute) Foundation, Yerevan, Armenia
\\
$^{2}$Benem\'{e}rita Universidad Aut\'{o}noma de Puebla, Puebla, Mexico
\\
$^{3}$Bogolyubov Institute for Theoretical Physics, Kiev, Ukraine
\\
$^{4}$Bose Institute, Department of Physics 
and Centre for Astroparticle Physics and Space Science (CAPSS), Kolkata, India
\\
$^{5}$Budker Institute for Nuclear Physics, Novosibirsk, Russia
\\
$^{6}$California Polytechnic State University, San Luis Obispo, California, United States
\\
$^{7}$Central China Normal University, Wuhan, China
\\
$^{8}$Centre de Calcul de l'IN2P3, Villeurbanne, Lyon, France
\\
$^{9}$Centro de Aplicaciones Tecnol\'{o}gicas y Desarrollo Nuclear (CEADEN), Havana, Cuba
\\
$^{10}$Centro de Investigaciones Energ\'{e}ticas Medioambientales y Tecnol\'{o}gicas (CIEMAT), Madrid, Spain
\\
$^{11}$Centro de Investigaci\'{o}n y de Estudios Avanzados (CINVESTAV), Mexico City and M\'{e}rida, Mexico
\\
$^{12}$Centro Fermi - Museo Storico della Fisica e Centro Studi e Ricerche ``Enrico Fermi', Rome, Italy
\\
$^{13}$Chicago State University, Chicago, Illinois, United States
\\
$^{14}$China Institute of Atomic Energy, Beijing, China
\\
$^{15}$COMSATS Institute of Information Technology (CIIT), Islamabad, Pakistan
\\
$^{16}$Departamento de F\'{\i}sica de Part\'{\i}culas and IGFAE, Universidad de Santiago de Compostela, Santiago de Compostela, Spain
\\
$^{17}$Department of Physics, Aligarh Muslim University, Aligarh, India
\\
$^{18}$Department of Physics, Ohio State University, Columbus, Ohio, United States
\\
$^{19}$Department of Physics, Sejong University, Seoul, South Korea
\\
$^{20}$Department of Physics, University of Oslo, Oslo, Norway
\\
$^{21}$Department of Physics and Technology, University of Bergen, Bergen, Norway
\\
$^{22}$Dipartimento di Fisica dell'Universit\`{a} 'La Sapienza'
and Sezione INFN, Rome, Italy
\\
$^{23}$Dipartimento di Fisica dell'Universit\`{a}
and Sezione INFN, Cagliari, Italy
\\
$^{24}$Dipartimento di Fisica dell'Universit\`{a}
and Sezione INFN, Trieste, Italy
\\
$^{25}$Dipartimento di Fisica dell'Universit\`{a}
and Sezione INFN, Turin, Italy
\\
$^{26}$Dipartimento di Fisica e Astronomia dell'Universit\`{a}
and Sezione INFN, Bologna, Italy
\\
$^{27}$Dipartimento di Fisica e Astronomia dell'Universit\`{a}
and Sezione INFN, Catania, Italy
\\
$^{28}$Dipartimento di Fisica e Astronomia dell'Universit\`{a}
and Sezione INFN, Padova, Italy
\\
$^{29}$Dipartimento di Fisica `E.R.~Caianiello' dell'Universit\`{a}
and Gruppo Collegato INFN, Salerno, Italy
\\
$^{30}$Dipartimento DISAT del Politecnico and Sezione INFN, Turin, Italy
\\
$^{31}$Dipartimento di Scienze e Innovazione Tecnologica dell'Universit\`{a} del Piemonte Orientale and INFN Sezione di Torino, Alessandria, Italy
\\
$^{32}$Dipartimento Interateneo di Fisica `M.~Merlin'
and Sezione INFN, Bari, Italy
\\
$^{33}$Division of Experimental High Energy Physics, University of Lund, Lund, Sweden
\\
$^{34}$European Organization for Nuclear Research (CERN), Geneva, Switzerland
\\
$^{35}$Excellence Cluster Universe, Technische Universit\"{a}t M\"{u}nchen, Munich, Germany
\\
$^{36}$Faculty of Engineering, Bergen University College, Bergen, Norway
\\
$^{37}$Faculty of Mathematics, Physics and Informatics, Comenius University, Bratislava, Slovakia
\\
$^{38}$Faculty of Nuclear Sciences and Physical Engineering, Czech Technical University in Prague, Prague, Czech Republic
\\
$^{39}$Faculty of Science, P.J.~\v{S}af\'{a}rik University, Ko\v{s}ice, Slovakia
\\
$^{40}$Faculty of Technology, Buskerud and Vestfold University College, Tonsberg, Norway
\\
$^{41}$Frankfurt Institute for Advanced Studies, Johann Wolfgang Goethe-Universit\"{a}t Frankfurt, Frankfurt, Germany
\\
$^{42}$Gangneung-Wonju National University, Gangneung, South Korea
\\
$^{43}$Gauhati University, Department of Physics, Guwahati, India
\\
$^{44}$Helmholtz-Institut f\"{u}r Strahlen- und Kernphysik, Rheinische Friedrich-Wilhelms-Universit\"{a}t Bonn, Bonn, Germany
\\
$^{45}$Helsinki Institute of Physics (HIP), Helsinki, Finland
\\
$^{46}$Hiroshima University, Hiroshima, Japan
\\
$^{47}$Indian Institute of Technology Bombay (IIT), Mumbai, India
\\
$^{48}$Indian Institute of Technology Indore, Indore, India
\\
$^{49}$Indonesian Institute of Sciences, Jakarta, Indonesia
\\
$^{50}$Inha University, Incheon, South Korea
\\
$^{51}$Institut de Physique Nucl\'eaire d'Orsay (IPNO), Universit\'e Paris-Sud, CNRS-IN2P3, Orsay, France
\\
$^{52}$Institute for Nuclear Research, Academy of Sciences, Moscow, Russia
\\
$^{53}$Institute for Subatomic Physics of Utrecht University, Utrecht, Netherlands
\\
$^{54}$Institute for Theoretical and Experimental Physics, Moscow, Russia
\\
$^{55}$Institute of Experimental Physics, Slovak Academy of Sciences, Ko\v{s}ice, Slovakia
\\
$^{56}$Institute of Physics, Academy of Sciences of the Czech Republic, Prague, Czech Republic
\\
$^{57}$Institute of Physics, Bhubaneswar, India
\\
$^{58}$Institute of Space Science (ISS), Bucharest, Romania
\\
$^{59}$Institut f\"{u}r Informatik, Johann Wolfgang Goethe-Universit\"{a}t Frankfurt, Frankfurt, Germany
\\
$^{60}$Institut f\"{u}r Kernphysik, Johann Wolfgang Goethe-Universit\"{a}t Frankfurt, Frankfurt, Germany
\\
$^{61}$Institut f\"{u}r Kernphysik, Westf\"{a}lische Wilhelms-Universit\"{a}t M\"{u}nster, M\"{u}nster, Germany
\\
$^{62}$Instituto de Ciencias Nucleares, Universidad Nacional Aut\'{o}noma de M\'{e}xico, Mexico City, Mexico
\\
$^{63}$Instituto de F\'{i}sica, Universidade Federal do Rio Grande do Sul (UFRGS), Porto Alegre, Brazil
\\
$^{64}$Instituto de F\'{\i}sica, Universidad Nacional Aut\'{o}noma de M\'{e}xico, Mexico City, Mexico
\\
$^{65}$IRFU, CEA, Universit\'{e} Paris-Saclay, F-91191 Gif-sur-Yvette, France, Saclay, France
\\
$^{66}$iThemba LABS, National Research Foundation, Somerset West, South Africa
\\
$^{67}$Joint Institute for Nuclear Research (JINR), Dubna, Russia
\\
$^{68}$Konkuk University, Seoul, South Korea
\\
$^{69}$Korea Institute of Science and Technology Information, Daejeon, South Korea
\\
$^{70}$KTO Karatay University, Konya, Turkey
\\
$^{71}$Laboratoire de Physique Corpusculaire (LPC), Clermont Universit\'{e}, Universit\'{e} Blaise Pascal, CNRS--IN2P3, Clermont-Ferrand, France
\\
$^{72}$Laboratoire de Physique Subatomique et de Cosmologie, Universit\'{e} Grenoble-Alpes, CNRS-IN2P3, Grenoble, France
\\
$^{73}$Laboratori Nazionali di Frascati, INFN, Frascati, Italy
\\
$^{74}$Laboratori Nazionali di Legnaro, INFN, Legnaro, Italy
\\
$^{75}$Lawrence Berkeley National Laboratory, Berkeley, California, United States
\\
$^{76}$Moscow Engineering Physics Institute, Moscow, Russia
\\
$^{77}$Nagasaki Institute of Applied Science, Nagasaki, Japan
\\
$^{78}$National and Kapodistrian University of Athens, Physics Department, Athens, Greece, Athens, Greece
\\
$^{79}$National Centre for Nuclear Studies, Warsaw, Poland
\\
$^{80}$National Institute for Physics and Nuclear Engineering, Bucharest, Romania
\\
$^{81}$National Institute of Science Education and Research, Bhubaneswar, India
\\
$^{82}$National Nuclear Research Center, Baku, Azerbaijan
\\
$^{83}$National Research Centre Kurchatov Institute, Moscow, Russia
\\
$^{84}$Niels Bohr Institute, University of Copenhagen, Copenhagen, Denmark
\\
$^{85}$Nikhef, Nationaal instituut voor subatomaire fysica, Amsterdam, Netherlands
\\
$^{86}$Nuclear Physics Group, STFC Daresbury Laboratory, Daresbury, United Kingdom
\\
$^{87}$Nuclear Physics Institute, Academy of Sciences of the Czech Republic, \v{R}e\v{z} u Prahy, Czech Republic
\\
$^{88}$Oak Ridge National Laboratory, Oak Ridge, Tennessee, United States
\\
$^{89}$Petersburg Nuclear Physics Institute, Gatchina, Russia
\\
$^{90}$Physics Department, Creighton University, Omaha, Nebraska, United States
\\
$^{91}$Physics Department, Panjab University, Chandigarh, India
\\
$^{92}$Physics Department, University of Cape Town, Cape Town, South Africa
\\
$^{93}$Physics Department, University of Jammu, Jammu, India
\\
$^{94}$Physics Department, University of Rajasthan, Jaipur, India
\\
$^{95}$Physikalisches Institut, Eberhard Karls Universit\"{a}t T\"{u}bingen, T\"{u}bingen, Germany
\\
$^{96}$Physikalisches Institut, Ruprecht-Karls-Universit\"{a}t Heidelberg, Heidelberg, Germany
\\
$^{97}$Physik Department, Technische Universit\"{a}t M\"{u}nchen, Munich, Germany
\\
$^{98}$Purdue University, West Lafayette, Indiana, United States
\\
$^{99}$Pusan National University, Pusan, South Korea
\\
$^{100}$Research Division and ExtreMe Matter Institute EMMI, GSI Helmholtzzentrum f\"ur Schwerionenforschung GmbH, Darmstadt, Germany
\\
$^{101}$Rudjer Bo\v{s}kovi\'{c} Institute, Zagreb, Croatia
\\
$^{102}$Russian Federal Nuclear Center (VNIIEF), Sarov, Russia
\\
$^{103}$Saha Institute of Nuclear Physics, Kolkata, India
\\
$^{104}$School of Physics and Astronomy, University of Birmingham, Birmingham, United Kingdom
\\
$^{105}$Secci\'{o}n F\'{\i}sica, Departamento de Ciencias, Pontificia Universidad Cat\'{o}lica del Per\'{u}, Lima, Peru
\\
$^{106}$Sezione INFN, Bari, Italy
\\
$^{107}$Sezione INFN, Bologna, Italy
\\
$^{108}$Sezione INFN, Cagliari, Italy
\\
$^{109}$Sezione INFN, Catania, Italy
\\
$^{110}$Sezione INFN, Padova, Italy
\\
$^{111}$Sezione INFN, Rome, Italy
\\
$^{112}$Sezione INFN, Trieste, Italy
\\
$^{113}$Sezione INFN, Turin, Italy
\\
$^{114}$SSC IHEP of NRC Kurchatov institute, Protvino, Russia
\\
$^{115}$Stefan Meyer Institut f\"{u}r Subatomare Physik (SMI), Vienna, Austria
\\
$^{116}$SUBATECH, Ecole des Mines de Nantes, Universit\'{e} de Nantes, CNRS-IN2P3, Nantes, France
\\
$^{117}$Suranaree University of Technology, Nakhon Ratchasima, Thailand
\\
$^{118}$Technical University of Ko\v{s}ice, Ko\v{s}ice, Slovakia
\\
$^{119}$Technical University of Split FESB, Split, Croatia
\\
$^{120}$The Henryk Niewodniczanski Institute of Nuclear Physics, Polish Academy of Sciences, Cracow, Poland
\\
$^{121}$The University of Texas at Austin, Physics Department, Austin, Texas, United States
\\
$^{122}$Universidad Aut\'{o}noma de Sinaloa, Culiac\'{a}n, Mexico
\\
$^{123}$Universidade de S\~{a}o Paulo (USP), S\~{a}o Paulo, Brazil
\\
$^{124}$Universidade Estadual de Campinas (UNICAMP), Campinas, Brazil
\\
$^{125}$Universidade Federal do ABC, Santo Andre, Brazil
\\
$^{126}$University of Houston, Houston, Texas, United States
\\
$^{127}$University of Jyv\"{a}skyl\"{a}, Jyv\"{a}skyl\"{a}, Finland
\\
$^{128}$University of Liverpool, Liverpool, United Kingdom
\\
$^{129}$University of Tennessee, Knoxville, Tennessee, United States
\\
$^{130}$University of the Witwatersrand, Johannesburg, South Africa
\\
$^{131}$University of Tokyo, Tokyo, Japan
\\
$^{132}$University of Tsukuba, Tsukuba, Japan
\\
$^{133}$University of Zagreb, Zagreb, Croatia
\\
$^{134}$Universit\'{e} de Lyon, Universit\'{e} Lyon 1, CNRS/IN2P3, IPN-Lyon, Villeurbanne, Lyon, France
\\
$^{135}$Universit\'{e} de Strasbourg, CNRS, IPHC UMR 7178, F-67000 Strasbourg, France, Strasbourg, France
\\
$^{136}$Universit\`{a} degli Studi di Pavia, Pavia, Italy
\\
$^{137}$Universit\`{a} di Brescia, Brescia, Italy
\\
$^{138}$V.~Fock Institute for Physics, St. Petersburg State University, St. Petersburg, Russia
\\
$^{139}$Variable Energy Cyclotron Centre, Kolkata, India
\\
$^{140}$Warsaw University of Technology, Warsaw, Poland
\\
$^{141}$Wayne State University, Detroit, Michigan, United States
\\
$^{142}$Wigner Research Centre for Physics, Hungarian Academy of Sciences, Budapest, Hungary
\\
$^{143}$Yale University, New Haven, Connecticut, United States
\\
$^{144}$Yonsei University, Seoul, South Korea
\\
$^{145}$Zentrum f\"{u}r Technologietransfer und Telekommunikation (ZTT), Fachhochschule Worms, Worms, Germany
\endgroup

  %%%%%%% done by webmaster team

\end{document}